\documentclass[useAMS,usenatbib]{mnras}
\usepackage{amsmath}
\usepackage{graphicx,txfonts,fleqn,enumerate,xcolor}

\newcommand*\justify{%
  \fontdimen2\font=0.4em
  \fontdimen3\font=0.2em
  \fontdimen4\font=0.1em
  \fontdimen7\font=0.1em
  \hyphenchar\font=`\-
}

\renewcommand*{\v}[1]  {\boldsymbol{#1}}

\newcommand*  {\twovector}[2] {{\begin{pmatrix} $1 \\ $2 \end{pmatrix}}}

\renewcommand {\emph}[1]  {\textit{#1}}

\newcommand   {\change}[1] {{\color{red}{#1}}}

\title[{\texttt{EnckeHH}: an integrator for gravitational dynamics}]
{{\texttt{EnckeHH}: an integrator for gravitational dynamics with a dominant mass that achieves optimal error behaviour}}

\author[David M. Hern{a}ndez and Matthew J. Holman]
	{David M. Hern{a}ndez$^{1,2}$\thanks{{{Email: dmhernandez@cfa.harvard.edu} }} and Matthew J. Holman$^{1}$ \\ 
	$^1$ Harvard--Smithsonian centre for Astrophysics, 60 Garden St., MS 51, Cambridge, MA 02138, USA \\
	}
\begin{document}

\maketitle

\label{first page}
\begin{abstract} 
 We present \texttt{EnckeHH}, a new, highly accurate code for orbital dynamics of perturbed Keplerian systems such as planetary systems or galactic centre systems.  It solves Encke's equations of motion, which assume perturbed Keplerian orbits.  By incorporating numerical techniques, we have made the code follow optimal roundoff error growth.  In a $10^ {12}$ day integration of the outer Solar System, \texttt{EnckeHH} was $3.5$ orders of magnitude more accurate than \texttt{IAS15} in a fixed time step test.  {Adaptive steps are recommended for \texttt{IAS15}.}  {  Through study of efficiency plots, we show that \texttt{EnckeHH} reaches significantly higher accuracy than the \texttt{Rebound} integrators \texttt{IAS15} and \texttt{WHCKL} for fixed step size. } 

\end{abstract}
\section{Introduction}
\label{sec:first}

The problem of predicting the orbital motions of planets has existed since before Newton wrote his universal law of gravitation.  Initially, such calculations were done by hand, but now we rely on computers to integrate the $n$-body equations of motion.  Generally,  there are two contexts for these computations.  In the first case, we are interested in determining precise trajectories over a time scale that is short compared to that of any relevant dynamical chaos \citep[e.g.,][]{Rein2019}.  We assume that the initial conditions and the physical forces are known well enough to support reliable predictions.  In the second, the goal is more often to integrate many different initial conditions to study long-term stability, explore the dynamical phase space, or to numerically test analytic theories.  For these purposes, speed is at a premium.  And it is essential to minimize numerical effects that might mimic physical effects that are not present in the actual system, such as dissipation.  There is some overlap, but the distinct demands of these two contexts has resulted in very different numerical algorithms.

The needs of solving for precise trajectories are often satisfied by generic conventional integrators for ordinary differential equations (ODEs).  Such integrators do not make assumptions about orbits, and require user input of an acceleration function, and sometimes higher  order time derivatives \citep[e.g.,][]{Makino1991}.  They can use adaptive steps sizes, which are particularly helpful in scale-free problems like gravity which result in stiff differential equations.  Several options exist, and their theory has been widely developed.  The Bulirsch--Stoer method \citep{press02,BulirschStoer1964} combines Richardson extrapolation with a simple one-step method like leapfrog.  Runge--Kutta methods (RK) are one-step methods that use substeps with predetermined spacings; examples are \texttt{RADAU} in \texttt{MERCURY6} \citep{C99,E85} and \texttt{IAS15} \citep{RS15}.  Multistep methods keep track of previous solution points to better make predictions; examples of these codes are described in \cite{QT90,Graz2005,Makino1991}.  Conventional integrators can be made particularly powerful if used in conjunction with well-behaved equations.  Regularization techniques remove the singularity of Newtonian gravity through changes of coordinates that result in more slowly varying orbits.  For computing near-Earth asteroid dynamics, regularization techniques like Kunstaanheimo--Stiefel and EDROMO \citep{SS71,Amato17,Bauetal2014} have been used.  The truncation error can be at the machine precision level or lower, making roundoff error the dominant error contribution; thus, some authors \citep{RS15,Graz2005,HMR07} have taken special care to control this error.  A curious feature about the methods described above, except for the regularization methods, is that they work for any ODEs, not just those describing planetary motion.  Arguably, the most popular of these currently is \texttt{IAS15} from \cite{RS15}.  

For the second case in which we need long-term calculations, conventional integrators will fail because they have truncation errors that make calculated orbits useless.   However, geometric, and especially symplectic  \citep{Ruth1983,chann90} integrators are suitable.  These methods conserve the Poincar\'{e} invariants of Hamiltonian dynamics. A limitation of symplectic integrators is they usually require a conservative Hamiltonian.  they can also require a very specific form of a Hamiltonian; any deviations from this form may require solving expensive implicit equations which defeat the purpose of constructing these methods for long-term calculations.  Their time step usually cannot be a function of the current state in contrast to conventional integrators.

Just as symplectic methods respect Poincar\'{e} invariants, other algorithms exploit different ``geometric'' properties of ODEs, resulting in powerful methods.  Examples include symmetric methods \citep{hair06,leim04,HB18}, which exploit the time-symmetry of some problems, projection methods \citep{hair06,Dengetal2020}, which force the solution to respect certain conserved quantities, and methods that respect the differentiability and smoothness of Hamiltonians \citep{H19,H19b}.

\cite{WH91} realized that using the averaging principle, which says that high frequency terms can be added or removed from the Hamiltonian governing planetary motion, can be used to write a symplectic map for planetary motion, which is widely known the Wisdom--Holman map (WH).  It forms the basis of many calculations of long-term planetary dynamics \citep{C99,DLL98,RT15,Reinetal2019,HD17}.  WH built upon a previous asteroidal map \citep{Wisdom82,W83}, itself a symplectic integrator (although not called that at the time).  WH assumes orbits are perturbations of Keplerian ellipses.  So its use is more limited in some ways than the aforementioned conventional methods.  For example, it will fail to accurately represent the motion during a close encounter of two planets. Some works have addressed some drawbacks of WH \citep{C99,TGS2015,Tamayoetal2019}. 

Ideally, we could use conventional integrators to solve equations that exploit the assumption of perturbed Keplerian motion, in the spirit of WH and regularization.  But actually, this task was achieved over $150$ years ago by Encke, who first wrote his eponymous equations of motion~ \citep{E1854}.  Arguably, Encke's method was more popular in the past, but it was largely replaced by methods that are simpler to program.  

In this work, we present a new, highly optimized code to solve Encke's equations, \texttt{EnckeHH}, where HH stands for Hernandez and Holman.  In our code, we have employed a Kepler advancer \citep{WH15} with excellent, simple error properties.  We have used various numerical techniques to further reduce its error.  We show our method follows optimal roundoff error growth, known as Brouwer's Law \citep{B37}; we also say the error is unbiased.  This is in contrast to \texttt{IAS15}, which shows biased error growth for fixed time steps.  {Fixed time steps are not recommended for \texttt{IAS15}.}  {An optimal fixed step code has many benefits; for example, we can use it if we know the position of a reference planet at fixed intervals.  The Encke method makes no assumptions about the form of the Hamiltonian, unlike symplectic composition methods.}  Unlike \texttt{IAS15}, we do not need to rely on adaptive stepping to mitigate error growth.  In an integration over $10^{12}$ days of the outer Solar System, \texttt{EnckeHH} had $3.5$ orders of magnitude smaller error than \texttt{IAS15} for the same step size.  We release our code to the public\footnote{\texttt{\justify github.com/matthewholman/rebound}}.  Instructions on using and running the code on a test problem are also available\footnote{\texttt{github.com/matthewholman/rebound/tree/master/examples/}\\{\texttt{encke\_hh/README.txt}}}.  Our paper is organized as follows.  Section \ref{sec:encke} presents the Encke equations of motion.  Section  \ref{sec:num} presents our optimizations to the method.  Section  \ref{sec:comp} shows short and long-term experiments demonstrating the performance of \texttt{EnckeHH}.  We conclude in Section \ref{sec:conc}.

\section{Encke Method}
\label{sec:encke}
We describe Encke's method; a reference is found in \cite{dan88}.  We assume a system with a dominant mass, the star, such that other bodies, the planets, are well described by perturbed Keplerian ellipses.  The planetary positions are written,
\begin{equation}
\label{eq:sum}
\v{x}_i = \v{\rho}_i + \v{\delta}_i,
\end{equation}
where $\v{\rho}_i$ is a Keplerian reference orbit and $ \v{\delta}_i$ constitutes the perturbations for planet $i$, and {$\v{x}_i$ and $\v{\rho}_i$ are heliocentric distances.}  The indices are in the range $i \in [1,N]$, where $N$ is the number of planets.  The index $i = 0$ is reserved for the star; its equations of motion are ignored in our code, so all indices in our paper are assumed greater than $0$.  

The perturbations can be due to any force besides the planet-star point mass Newtonian gravity.  The differential equations describing the reference orbits are,
\begin{equation}
\label{eq:ref}
\ddot{\v{\rho}}_i= -\mu_i \frac{\v{\rho}_i}{{\rho}_{i}^3},
\end{equation}
where $\mu_i = G (m_0 + m_i)$, $G$ is the gravitational constant, and $m_0$ and $m_i$ are the star and planet mass, respectively.  The positions $\v{x}_i$ can be found from integrating,
\begin{equation}
\label{eq:real}
\ddot{\v{x}}_{i} = -\mu_i \frac{\v{x}_i}{{x}_i^3} + \v{a}_i(\v{x},\dot{\v{x}},t),
\end{equation}
where $\v{a}_i$ are accelerations which may depend on the time $t$.  The $\v{a}_i$ can cause the set of ODEs \eqref{eq:real} to be coupled.  Eqs. \eqref{eq:ref} and \eqref{eq:real} are each initial value problems and require $6N$ initial conditions, with $N$ the number of planets.

We can write an equation for $\v{\delta}_i$ from the difference of eqs. \eqref{eq:ref} and \eqref{eq:real}:
\begin{equation}
\label{eq:encke}
\ddot{\v{\delta}_i} = \mu_i \left(\frac{\v{\rho}_i}{\rho_i^3} - \frac{\v{x}_i}{{x}_i^3}\right) + \v{a}_i(\v{x},\dot{\v{x}},t).
\end{equation}
The quantity in large parentheses is numerically problematic, involving a difference of nearly equal numbers.  A useful form is given by,
\begin{equation}
\label{eq:encke2}
\ddot{\v{\delta}_i} = - \frac{\mu_i}{\rho_i^3} \v{\delta}_i + \frac{\mu_i}{{\rho}_i^3} \left(1 -  \frac{\rho_i^3} {{x}_i^3} \right) \v{x}_i  + \v{a}_i(\v{x},\dot{\v{x}},t).
\end{equation}
As written, the quantity in parentheses is still problematic, but it can be recast in a form that avoids the difference of similar quantities~\citep{Battin.1987}.  We first define the auxiliary quantity,
\begin{equation}
\label{eq:q}
q_i = \frac{\left( \v{\delta}_i + 2 \v{\rho}_i\right) \cdot \v{\delta}_i}{{\rho}_i^2}.
\end{equation}
There are other forms of $q_i$ in the literature, but we found this one appealing because it avoids the use of $\v{x}_i$, whose calculation is prone to roundoff error (see Section \ref{sec:num}). 
Then we define the function,
\begin{equation}
\label{eq:fq}
f(q_i) = \left(1 -  \frac{\rho_i^3} {{x}_i^3} \right),
\end{equation}
which can be rewritten as,
\begin{equation}
\label{eq:f}
f(q_i) = q_i \frac{3 + 3 q_i + q_i^2}{(1 + q_i)^{3/2} + (1 + q_i)^3}.
\end{equation}
Form \eqref{eq:f} is the one we use.  It can be represented as a series in the small parameter $q_i$, but we opted for this exact representation.  As a demonstration of the importance of properly treating this difference of like quantities, let $\v{\rho}_i = (1.0,0,0)$ and $\v{\delta}_i = (10^{-14},0,0)$.  On a machine using double precision arithmetic in IEEE standard, form \eqref{eq:fq} gives $f(q_i) = 2.997602166487923\times 10^{-14}$ while form \eqref{eq:f} gives a more correct $f(q_i) = 2.999999999999941\times 10^{-14}$.

The final equations of motion are,
\begin{equation}
\label{eq:final}
\ddot{\v{\delta}_i}=  - \frac{\mu_i}{\rho_i^3} \left[ \v{\delta}_i  - f(q_i) \v{x}_i\right]  + \v{a}_i(\v{x},\dot{\v{x}},t),
\end{equation}
which we call Encke's equations of motion.

\subsection{Calculation of accelerations}
\label{sec:accel}
As an example for the form of $\v{a}_i(\v{x},\dot{\v{x}},t)$, assume it is due to point mass contributions from the other planets.  Let $\v{x}_{i j} = \v{x}_i - \v{x}_j$.  Then it takes form,
\begin{equation}
\label{eq:acc}
\v{a}_i(\v{x}) = -Gm_j \sum_{j \ne i} \left(\frac{\v{x}_{i j}}{{x}_{i j}^3} + \frac{\v{x}_{j}}{{x}_{j}^3} \right).
\end{equation}
Immediately, the $3N$ equations \eqref{eq:final} have become coupled.  This form of the acceleration is assumed in the remainder of this paper.

\section{Numerical Implementation}
\label{sec:num}
\subsection{Solving the Kepler equation}
\label{sec:Kep}
To implement our Encke method, we need to solve both eqs. \eqref{eq:ref} and \eqref{eq:final}.  Eq. \eqref{eq:ref} can be solved as a function of time, and \eqref{eq:final} depends on this solution.  First, we focus on solving the initial value problem \eqref{eq:ref}.  Fortunately, there exist excellent modern methods to solve this such as those in \cite{RT15,WH15}.  The \cite{WH15} solver, \texttt{Universal.c}, avoids the Strumpff series expansion used by \cite{RT15}, in the same way we avoid the series expansion of \eqref{eq:f}, and in this paper we use a modified \texttt{Universal.c}.  Our modification is as follows.

When solving for the universal variable $s$ in eq. (35) of \cite{WH15} we have iterated until convergence.  Let $s^{(k)}$ be the value of $s$ after $k$ iterations.  Our criteria for convergence is that either $s^{(k)} = s^{(k-1)} $ or $s^{(k)} = s^{(k-2)}$.  For the iteration strategy we use a Newton method first.  A failsafe is provided by the bisection method.  Finally, the timestep is divided if both of these fail.  Otherwise, \texttt{Universal.c} remains the same.  In a typical planetary system problem, bisection is rarely used.  For example, when integrating the outer Solar System for $10^{7}$ days using $h = 100$ days, where $h$ is the stepsize, the bisection method was only used $4$ times in $8 \times 10^{5}$ calls to the Kepler advancer.

Any update to a reference orbit from the Kepler solver used compensated summation \citep{Kahan1965,hair06,Wisdom2018} to mitigate roundoff error.  Compensated summation solves the problem of the precision lost when adding two numbers that are significantly different in order of magnitude by recording the error made after each sum and later compensating for the error.
\subsection{Solving the Encke equations of motion}

Solving the Encke equations of motion, \eqref{eq:final}, requires an ODE solver.  We have opted for the power series method used by works such as \cite{E85} and \cite{RS15}.  As shown by \cite{E74}, this series method is equivalent to a Runge--Kutta (RK) method, so in what follows, we will also refer to it as an RK method.  This particular RK method is 15th order and uses Gauss--Radau spacings for the substeps.  We have built upon code from the \cite{RS15} implementation, \texttt{IAS15}.  As in earlier works, we used fixed-point iteration to solve for the RK implicit set of equations,  with an initial guess obtained from extrapolation.

As an iteration convergence criteria for the RK method, we retained the strict error tolerance from \texttt{IAS15}, which uses the last term in the power series for the acceleration.  We have attempted various less strict criteria, which did not work as well, despite ${\delta_i}$ usually being small.  As an example, we tried enforcing convergence in $\v{x}_i$ rather than $\v{\delta}_i$.  In addition, we allow iterations to stop if the error has grown compared to a previous iteration.  This situation occurred rarely.  In the same test as in Sec. \ref{sec:Kep}, it occurred in $248$ steps, or $0.3\%$ of the timesteps. 

We have taken care to reduce roundoff error by implementing various numerical techniques--- for example, compensated summation again.  The additions in \eqref{eq:sum}, \eqref{eq:final}, and \eqref{eq:acc} have been performed with compensated summation.  

We have used rectification, which is resetting of the reference orbit, once ${\delta}_i$ becomes too large.  Specifically, for the rectification criterion we have used,
\begin{equation}
\label{eq:rect}
\Delta_i = \frac{\delta_i}{a_i (1 - e_i)}  > \epsilon,
\end{equation}
where $a_i$ and $e_i$ are the semi-major axis and eccentricity of the reference orbit, respectively.  $\epsilon$ is specified by the user, and we set it by default to $0.01$.  This value was used in tests in this paper.    The criteria \eqref{eq:rect} is not perfect; in particular, a concern is that, especially for large steps, $\delta_i$ may become large in the middle of the timestep.  However, this was never a problem in our tests.  For example, running our Solar System problem for $10^8$ days with stepsizes from $0$ to $400$ days, the condition $|(\Delta_i - \epsilon)/\epsilon | > 1$ was never satisfied.  Using $h = 40$ days, a rectification was performed approximately every $139$ steps.

The new orbit is defined by, 
\begin{equation}
\begin{aligned}
\v{\rho}_{i}^{\prime} &=& \v{\rho}_{i} + \v{\delta}_i,& \\
\dot{\v{\rho}}_{i}^{\prime} &=& \dot{\v{\rho}}_{i} + \dot{\v{\delta}}_i,& \\
\v{\delta}_i^\prime &=& 0, &~~~~\mathrm{and}   \\
\dot{\v{\delta}}_i^\prime &=& 0,& 
\end{aligned}
\end{equation}
where primes indicate new values.  If $\epsilon$ is too large, the reference orbit is no longer a good approximation and the assumption of $\v{\delta}_i$ small, made in \texttt{EnckeHH}, breaks down.  {In addition, biases from the RK method could be seen if $\epsilon$ is made large, although we have not seen this in our experiments.}  If $\epsilon$ is too small, frequent rectifications occur which slow the code down and increase the number of iterations the RK method needs to converge.  This is because the initial guess for the accelerations at the substeps, from extrapolation at the previous step, can no longer be used.  For our value of $\epsilon$, this was never a problem.  Running the outer Solar System for $10^7$ days with $h = 40$ days, we calculate the average number of iterations after a rectification is done, and compare it to the average otherwise.  These values are $4.3$ and $2.4$, respectively.  For $h = 400$ days, these values are $7.2$ and $6.6$, respectively.  These differences in iteration number affected computation times in our tests negligibly.

\subsection{Other numerical considerations}

When designing \texttt{EnckeHH}, there were a number of numerical issues we considered.  An incomplete list is given here:
\begin{enumerate}
\item Our choice of ODE solver for eq. \eqref{eq:final} can be improved in future work.  For example, this method uses fixed-point iteration to converge to the correct accelerations at $7$ substeps, and such an approach requires many iterations for larger steps.  However, using a multidimensional Newton iteration \citep{Gonzalezpinto01} allows for larger step sizes, which is the strength of the Encke method.  

\item Using Gauss--Radau spacings has advantages described in \cite{E74}.  However, a disadvantage is that they lead to a mapping that is neither symmetric nor symplectic.  This is in contrast to Gauss--Lobatto spacings, which give a symmetric map, or Gauss--Legendre spacings, which lead to a symmetric and symplectic map \citep{hair06,FHP04}.  If the truncation error is always guaranteed to be less than the machine precision, the geometric properties of maps with these spacings becomes irrelevant.  But if this guarantee is broken for even a tiny fraction of the time, error biases may be introduced.  The use of symplectic spacings allowed simplification of the error analysis in \cite{HMR07}.   

\item \cite{RS15,HMR07} advocated for defining constants as accurately as possible, but we have not found evidence for special care taken in defining constants in \texttt{IAS15},  and we have not explored this issue ourselves.

\item Adaptive stepping mitigates roundoff error biases to some extent, as correctly claimed by \cite{RS15}.  However, to make a code as unbiased as possible, we focus on behaviour using fixed timestep.  Adaptive timestepping may not be practical for all problems, for instance, if the information of a reference body is known only at specific time intervals.  In addition, adaptive stepping is not a perfect solution for biased errors.  Next, there is no reason adaptive stepping should be needed for planetary problems with well-defined timescales that we focus on here, except for extremely eccentric orbits for which $e_i \approx 1$.  For such orbits, we have found an adaptive stepping routine useful, which we include in \texttt{EnckeHH}, but we do not focus on it here.  Finally, one might argue that introducing an unphysical parameter $\epsilon$ to control the adaptive stepsize is unsatisfying.  \cite{RS15} argue this parameter should be about $\epsilon = 0.01$ but use a default of $\epsilon = 10^{-9}$. 

\end{enumerate}

\section{Numerical experiments}
\label{sec:comp}

With \texttt{EnckeHH}, we have been able to produce the most accurate integrations of the outer Solar System {in our comparisons.} 

\subsection{Short-term simulations}
\label{sec:short}
Our test problem is the Sun plus four outer giant planets.  Initial conditions from \cite{Applegate1986} are used.  The system of units is au, days, and solar mass.  We first run short-term integrations for $5\times 10^{8}$ days $\approx 115,423 P$, where $P$ is the Jupiter period, and compare \texttt{EnckeHH} with \texttt{IAS15}.  {We use `short-term' loosely, meaning a small number of Lyapunov times.  The Lyapunov time of the Jovian planet system is not well known \citep{Hayes2007}}.  We use a constant time step, which turns out to be a challenging test.  We compare against the \texttt{Rebound} integrator, \texttt{IAS15}.  We use the \texttt{Rebound} options, \\
\texttt{r->heartbeat = heartbeat;} \\
\texttt{r->force\_is\_velocity\_dependent = 0;} \\
\texttt{        r->integrator    = REB\_INTEGRATOR\_IAS15;} \\
\texttt{        r->gravity    = REB\_GRAVITY\_COMPENSATED;} \\
\texttt{        r->ri\_ias15.epsilon         = 0.; }. \\
Note, we have turned on compensated gravity for \texttt{IAS15} in an attempt to make the \texttt{IAS15} integration as accurate as possible (compensated summation is always used in \texttt{EnckeHH}).  {We also compare against the \texttt{Rebound} integrator \texttt{WHCKL}.  In Section 4 of \cite{Reinetal2019b}, it is stated that \texttt{WHCKL} is their fastest integrator for their outer Solar System test, so we choose it here.}  {\texttt{WHCKL} is symplectic and not amenable to time-dependent Hamiltonians, unlike \texttt{IAS15} or \texttt{EnckeHH}.  Nonetheless, we can use it to study the basic outer Solar system.  To use this method, we apply the \texttt{Rebound} options, \\
\texttt{r->integrator           = REB\_INTEGRATOR\_WHFAST;} \\
\texttt{r->ri\_whfast.safe\_mode  = 0;}    \\   
\texttt{r->ri\_whfast.corrector  = 17;}      \\
\texttt{r->ri\_whfast.kernel     = REB\_WHFAST\_KERNEL\_LAZY; }  \\
\texttt{r->heartbeat = heartbeat;} \\
\texttt{r->force\_is\_velocity\_dependent = 0;} \\
\texttt{        r->gravity    = REB\_GRAVITY\_COMPENSATED;} \\}
{We have used these options to make \texttt{WHCKL} as fast and accurate as possible.}  

\begin{figure}
	\includegraphics[width=90mm]{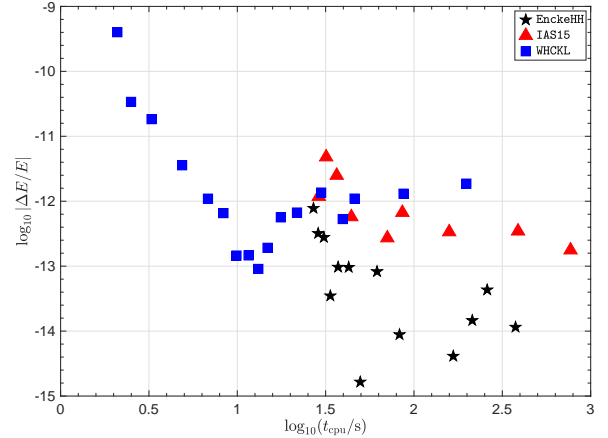}
	\caption{Energy error vs cpu time for short-term outer Solar System integrations over $5 \times 10^8$ days. {\texttt{IAS15} and \texttt{WHCKL} do not reach \texttt{EnckeHH}'s accuracy.}  
	\label{fig:eff1}
  	}
\end{figure}

\begin{figure}
	\includegraphics[width=90mm]{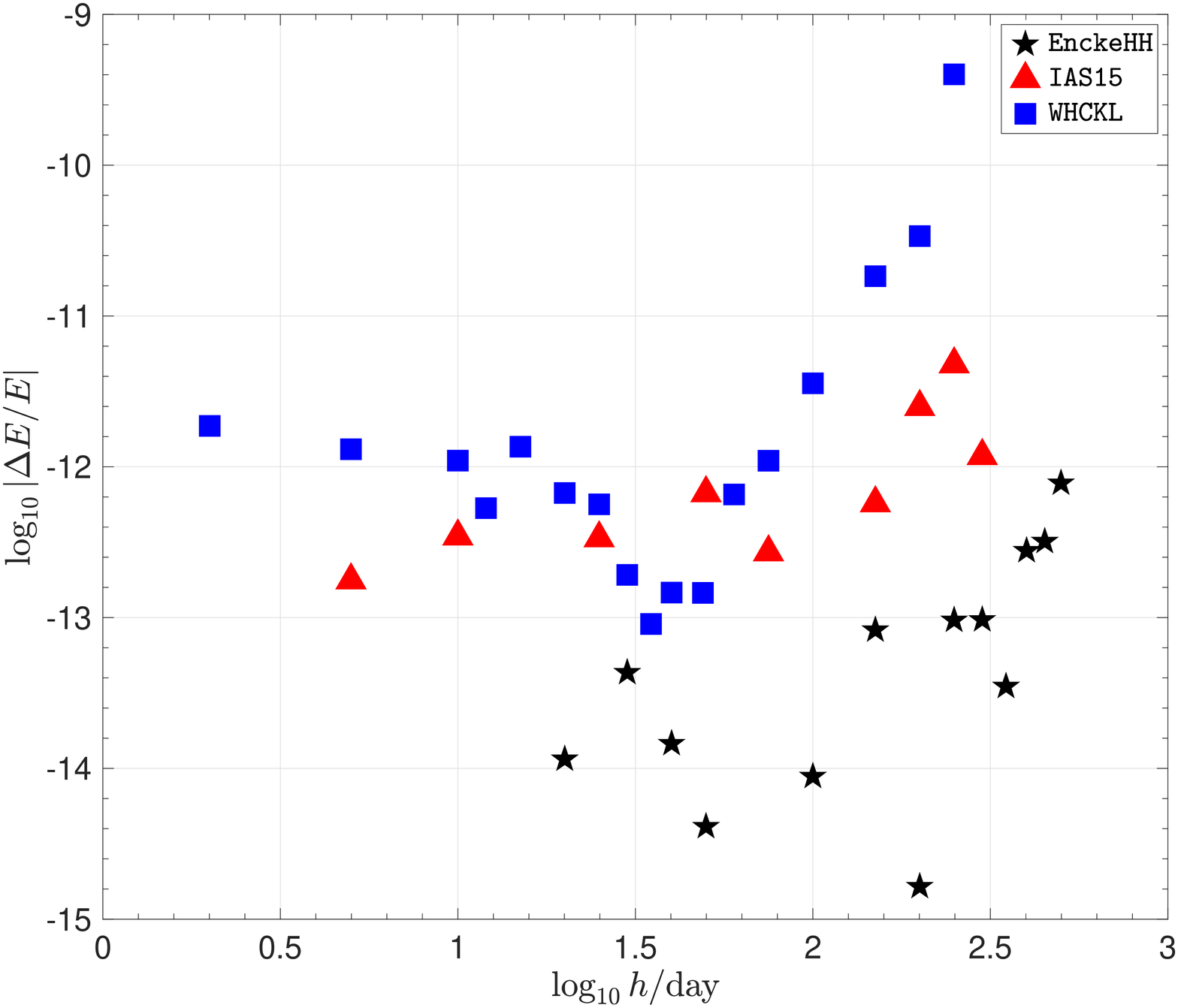}
	\caption{Energy error vs time step for short-term outer Solar System integrations over $5 \times 10^8$ days.  {For a given step size, \texttt{EnckeHH} is always the most accurate method and achieves a small error over a large range of step sizes.}
	\label{fig:eff2}
  	}
\end{figure}

{In Fig. \ref{fig:eff1}, we plot computing time\footnote{{When producing Fig. \ref{fig:eff1}, we used a non public version of \texttt{EnckeHH}, which gives identical results, but is slower on average by $28\%$ for $5$ timesteps.  We have thus corrected the computing times of Fig. \ref{fig:eff1} to estimate those of \texttt{EnckeHH}, although the change to this plot is minimal.}} versus energy error for the integrators.  We see \texttt{EnckeHH} reaches accuracies the other methods do not.  The computing time of \texttt{EnckeHH} is somewhat large, due to the expensive Kepler solver, which uses similar computing resources as the acceleration calculation.  In Fig. \ref{fig:eff2}, we plot the error versus timestep.  For a narrow range of timesteps between $30$ and $60$ days, the \texttt{WHCKL}  error improves.  However, \texttt{EnckeHH} has a stable, small error over more than an order magnitude change in timesteps.  We also see that for the same timestep, the error of \texttt{EnckeHH} is always smallest.  }

{Judging from the plots of \cite{Reinetal2019b}, it was reasonable to expect \texttt{WHCKL} would not match \texttt{EnckeHH}'s high accuracies.  Related discussion is in \cite{Hetal2020}.  A disadvantage of the higher order methods of \cite{Reinetal2019b} is that each step is made up of a composition of operators, possibly exacerbating roundoff errors.  }

In the remaining tests in this paper, we will continue to use the energy error as a metric for how ``good'' \texttt{EnckeHH} is.  We justify this approach by showing the accuracy in phase space for the test problem of Figs. \ref{fig:eff1} and \ref{fig:eff2}.  To measure the accuracy in phase space, we use a reversibility test.  Unlike methods like WH, \texttt{EnckeHH} is not formally time-reversible/symmetric \citep{hair06,HB18}.  If we integrate the problem of Fig. \ref{fig:eff1} and \ref{fig:eff2} with WH with any step size, reverse the sign of velocities, and integrate the same number of steps, WH would exactly recover the initial conditions on an infinitely precise machine, despite possibly large errors in the energy.  The same is not true of \texttt{EnckeHH}: if it gives a time-reversible solution, its only because it has solved the phase space from the equations of motion \eqref{eq:final} exactly, and thus the energy error is also $0$.  

In our test, we integrate forward the problem of Figs. \ref{fig:eff1} and \ref{fig:eff2}, but only for $5 \times 10^7$ days, reverse the sign of the time step, and integrate backward.  We then calculate the $L_2$ norm, defined as,
\begin{equation}
\label{eq:L2}
L_2 = \sqrt{\sum_i (\v{x}_i - \v{x}_i^\prime )^2 },
\end{equation}
where $\v{x}_i$ are the initial planetary positions, and $\v{x}_i^\prime$ are the final coordinates.  In Table \ref{tab:l2}, we show the value of $L_2$ and the energy error for various step sizes.  The energy error is calculated after the forward integration.  We see the energy error is an approximate proxy for $L_2$ {in the sense that they grow at approximately similar rates.  The $L_2$ norm and absolute value of the energy error grow by $1.0\times10^{-4}$ and $4.0\times10^{-5}$, respectively, as we increase the stepsize.}  This justifies our use of energy error as a measure of the ``goodness'' of \texttt{EnckeHH}.  Further work on the relationship between energy error and phase space accuracy was done in \cite{Hetal2020}.  {After many Lyapunov times, we expect a reversibility measure to be poor, while the energy error can still remain small.}  
\begin{table}
\label{tab:l2}
\begin{tabular}{lll}
$h$   &  $|\Delta E/E| $ & $L_2$            \\
$100$  & $1.9 \times 10^{-15}$ & $1.2\times 10^{-8}$ \\
$300$ & $1.4 \times 10^{-15}$ & $1.5\times 10^{-8}$\\
$500$ & $8.8 \times 10^{-15}$ & $3.5\times 10^{-7}$\\
$700$ & $6.4 \times 10^{-12}$ & $4.6\times 10^{-6}$\\
$900$ & $4.8 \times 10^{-11}$ & $1.2\times 10^{-4}$
\end{tabular}
\caption{{Results of the reversibility test of \texttt{EnckeHH} applied to the test problem of Figs. \ref{fig:eff1} and \ref{fig:eff2}.  The energy error is a rough proxy for the phase space error.}   }
\end{table}

\begin{figure*}
	\includegraphics[width=180mm]{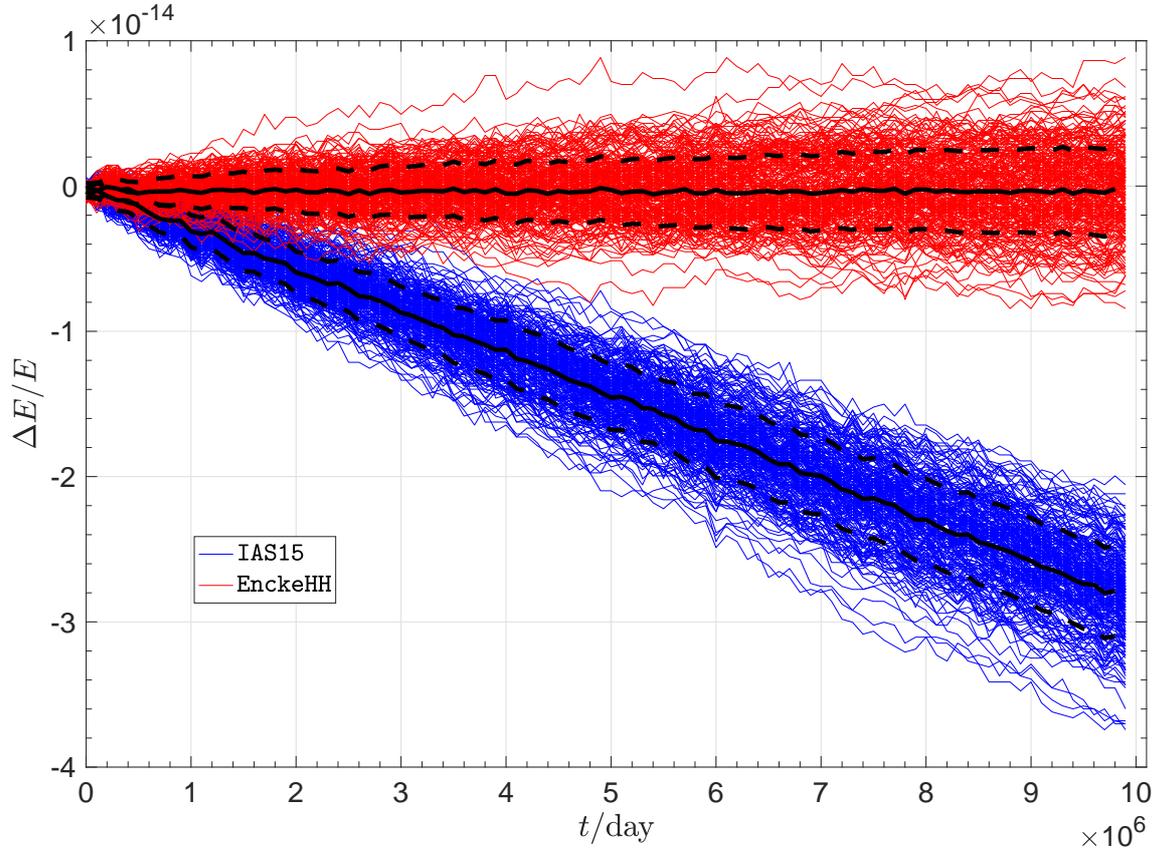}
	\caption{Energy error vs time for $125$ outer Solar System initial conditions integrated over $10^7$ days, and using $h = 40$ days.  The mean\change{s} and standard deviations have been plotted for $1000$ initial conditions. The \texttt{IAS15} integrations show linear bias while bias has been largely suppressed in the \texttt{EnckeHH} data.
	\label{fig:1000sample}
  	}
\end{figure*}

Next, we examine in Fig. \ref{fig:1000sample} the statistical error over time of $1000$ nearby initial conditions.  The outer Solar System is run for $10^7$ days with $h = 40$ days.  The \texttt{IAS15} options remain the same.  The initial conditions are generated by displacing Jupiter's $x$-position by an amount,
\begin{equation}
\label{eq:perturb}
\eta = n\times 10^{-14},
\end{equation}
where $n$ is an integer such that $n \in [0,1000)$.  Only $125$ initial conditions with each method are plotted for clarity of the figure.  $100$ points are used to plot each curve.  The means are indicated with a thick line, while dashed lines indicate sample standard deviations.  It is clear already in this short-term integration that \texttt{IAS15} suffers from a linear bias in its roundoff error.  \cite{HMR07} suggest that a linear bias can result from constants that are not carefully defined, so this may be an explanation.  The error of \texttt{EnckeHH} is dominated by Kepler solver error, which has been investigated carefully and reduced previously \citep{WH15}.  The maximum value of the absolute value of the mean for \texttt{EnckeHH} is $5.8 \times 10^{-16} = 2.6 \alpha$, where $\alpha$ is the machine precision, which may indicate a nearly negligible bias.  The sample standard deviation follows Brouwer's Law  and is described by, $s = 5.7 \times 10^{-18} \sqrt{n} + 2.1\times 10^{-16}$, where $n = t/h$, and $t = 10^{7}$ days.  Thus, the error per timestep is $\approx 0.01 \alpha$, so roundoff errors have been clearly suppressed as the errors are committed.

\begin{figure}
	\includegraphics[width=90mm]{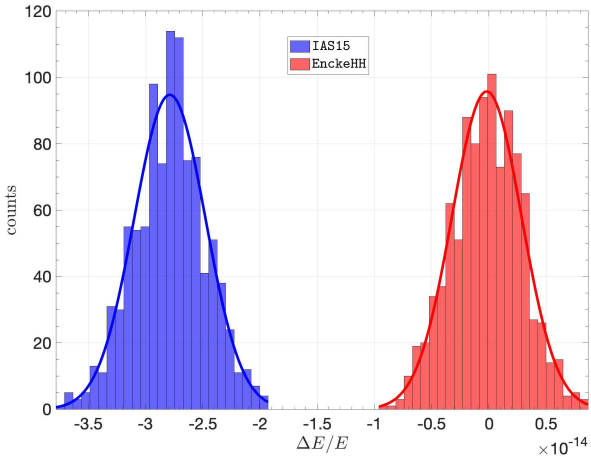}
	\caption{	Histogram of the energy errors at the last step of $1000$ initial conditions of the outer Solar System, integrated over $10^7$ days.  The mean and standard deviations in the last time step of Fig. \ref{fig:1000sample} have been used to compute normal curves, which are also plotted.  The data are approximately normally distributed, but only the \texttt{EnckeHH} data are approximately centred around 0.  More specifically, the \texttt{IAS15} and \texttt{EnckeHH} means are $-2.8\times 10^{-14}$ and $-1.9\times 10^{-16} $, respectively, while the standard deviations are $3.12 \times 10^{-15}$ and $3.04 \times 10^{-15} $.
	\label{fig:hist}
  	}
\end{figure}

To investigate the statistics further, Fig. \ref{fig:hist} plots a histogram of the errors at the last timestep.  Normal distributions with the final means and standard deviations have been plotted as well.  The errors for both \texttt{EnckeHH} and \texttt{IAS15} approximately follow normal distributions with sample standard deviations of $3.04 \times 10^{-15}$ and $3.12\times 10^{-15}$, respectively.  By calculating the error in these sample standard deviations, we show in Appendix \ref{sec:std} that they are consistent.  

\subsection{Long-term simulations}

\begin{figure*}
	\includegraphics[width=180mm]{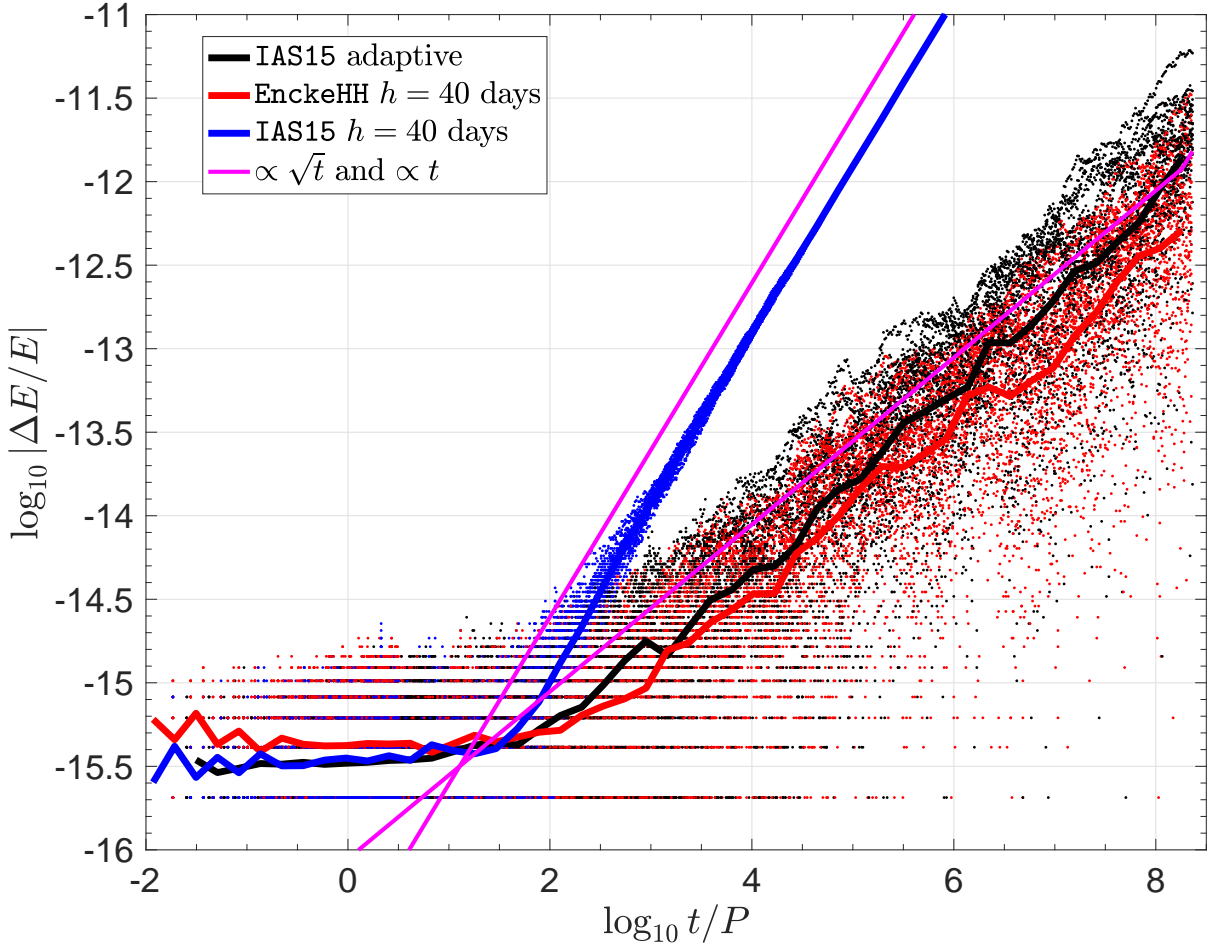}
	\caption{$20$ integrations from Fig. \ref{fig:1000sample} are extended to $10^{12}$ days.  We have also used \texttt{IAS15} with its recommended adaptive stepping parameter $\epsilon=10^{-9}$.  $1000$ points are plotted per integration.  A mean in log space is indicated with thick lines.  The linear bias in the fixed step \texttt{IAS15} case persists.  Its bias is mitigated, but not eliminated, when adaptive stepping is used.  The error for \texttt{EnckeHH} usually remains below the \texttt{IAS15} error. 
	\label{fig:longt}
  	}
\end{figure*}

Finally, we study long-term simulations of the outer Solar System over $10^{12}$ days $\approx 2.3\times 10^8 P$, in Fig. \ref{fig:longt}, where the log of the absolute error is plotted versus time.  The same timesteps as in Section \ref{sec:short} are used for \texttt{EnckeHH} and \texttt{IAS15}, but we have also calculated $h = 200$ runs with \texttt{EnckeHH}, not shown here, and its roundoff error grew at the same rate.  We also plot \texttt{IAS15} using adaptive stepping as recommended in \cite{RS15}, using a starting timestep of $40$ days, and adaptive step option,\\
\texttt{        r->ri\_ias15.epsilon         = 1e-9}. \\
For a run over $10^8$ days, this adaptive stepping yields an average stepsize of $h = 85$ days.  For each of the three methods, $1000$ logarithmically spaced outputs per run have been plotted and $20$ runs have been performed, by generating initial conditions according to eq. \eqref{eq:perturb} with $n\in [0,20)$.  Mean errors in log space are indicated with solid lines.  

Again, a clear linear bias is seen in the \texttt{IAS15} fixed step case.  Adaptive stepping mitigates, but does not eliminate completely, the biased behaviour.  {The slope in its median curve from $\log_{10} (t/P) = 2.1$ onwards is $0.54$ while the slope for \texttt{EnckeHH} is $0.49$.}  The \texttt{EnckeHH} behaviour {is} unbiased with an error that usually remains below that of \texttt{IAS15}, even with adaptive stepping.  At the end of the integration, the mean \texttt{EnckeHH} error is $10^{-3.5}$ times that of the \texttt{IAS15} error with fixed step.  {Because of the non-zero bias of the \texttt{IAS15} variable step method, we expect the advantage of \texttt{EnckeHH} over it to grow in time in this test.  For $h = 40$ days, \texttt{EnckeHH} was more accurate than \texttt{IAS15} with our chosen adaptive step option.  Of course, it will not be more accurate than \texttt{IAS15} for all steps.}

\section{Conclusions}
\label{sec:conc}

We have presented a code for calculating planetary dynamics orbits with high accuracy.  Our code has unbiased roundoff error growth, following Brouwer's Law.  \texttt{EnckeHH} is significantly more accurate than the code \texttt{IAS15} for fixed timesteps: in integrations of the outer Solar System over $10^{12}$ days, \texttt{EnckeHH} was $3.5$ orders of magnitude more accurate than \texttt{IAS15} for a timestep.  We have improved the error behaviour of \texttt{EnckeHH} by using numerical tools such as compensated summation and by being careful with the convergence of the solution of implicit equations, such as the Kepler equation.

While other codes like \texttt{IAS15} solve generic ODEs, \texttt{EnckeHH} is optimized for planetary orbits and, indeed, any gravitational $N$-body problem with a dominant central mass.  It does not rely on adaptive stepping like \texttt{IAS15} to mitigate its roundoff error.  

{Through exploration of efficiency plots, we show \texttt{IAS15} and \texttt{WHCKL}, also in \texttt{Rebound}, do not reach \texttt{EnckeHH}'s accuracy for fixed time step tests.}  Our code allows for adaptive stepping, but we did not find this necessary except for orbits that were nearly parabolic.  

Many further improvements to \texttt{EnckeHH} are possible.  Examples include:
\begin{enumerate}
\item changing the iteration scheme from fixed point to Newtonian,
\item using different RK spacings with better error properties,
\item modifying the code so that arbitrary forces can be input, and 
\item making the code able to handle close encounters by switching to a generic ODE solver when some criteria is satisfied. 
\end{enumerate}

\section{Acknowledgements}
We thank Sam Hadden and Matthew Payne for helpful discussions, and Daniel Tamayo for a helpful referee report.

\section{Data Availability}
Our code is released to the public. Instructions on how to use \texttt{EnckeHH} on a test problem are available\footnote{\texttt{github.com/matthewholman/rebound/tree/master/}\\ \texttt{examples/encke\_hh/README.txt}}.

\appendix
\section{Error in the sample standard deviation}
\label{sec:std}

We compute the error in the sample standard deviations, $s$, reported in Section \ref{sec:short}.  For large sample sizes, an unbiased estimator of this error, $\sigma$, is found from,
\begin{equation}
\sigma = s \sqrt{\mathrm{e}\cdot \left(1 - \frac{1}{n}\right)^{n-1} - 1},
\end{equation}
where $n=1000$ is the sample size.  Using the $s$ from Section \ref{sec:short}, we find $\sigma = 6.8 \times 10^{-17}$ and  $\sigma = 7.0 \times 10^{-17}$ for \texttt{EnckeHH} and \texttt{IAS15}, respectively.  This implies the two $s$ are consistent.
\bibliographystyle{mnras}
\bibliography{paper}

\begin{thebibliography}{}
\makeatletter
\relax
\def\mn@urlcharsother{\let\do\@makeother \do\$\do\&\do\#\do\^\do\_\do\%\do\~}
\def\mn@doi{\begingroup\mn@urlcharsother \@ifnextchar [ {\mn@doi@}
  {\mn@doi@[]}}
\def\mn@doi@[#1]#2{\def\@tempa{#1}\ifx\@tempa\@empty \href
  {http://dx.doi.org/#2} {doi:#2}\else \href {http://dx.doi.org/#2} {#1}\fi
  \endgroup}
\def\mn@eprint#1#2{\mn@eprint@#1:#2::\@nil}
\def\mn@eprint@arXiv#1{\href {http://arxiv.org/abs/#1} {{\tt arXiv:#1}}}
\def\mn@eprint@dblp#1{\href {http://dblp.uni-trier.de/rec/bibtex/#1.xml}
  {dblp:#1}}
\def\mn@eprint@#1:#2:#3:#4\@nil{\def\@tempa {#1}\def\@tempb {#2}\def\@tempc
  {#3}\ifx \@tempc \@empty \let \@tempc \@tempb \let \@tempb \@tempa \fi \ifx
  \@tempb \@empty \def\@tempb {arXiv}\fi \@ifundefined
  {mn@eprint@\@tempb}{\@tempb:\@tempc}{\expandafter \expandafter \csname
  mn@eprint@\@tempb\endcsname \expandafter{\@tempc}}}

\bibitem[\protect\citeauthoryear{{Amato}, {Bau}  \& {Bombardelli}}{{Amato}
  et~al.}{2017}]{Amato17}
{Amato} D.,  {Bau} G.,   {Bombardelli} C.,  2017, Monthly Notices of the Royal
  Astronomical Society, 470, 2079

\bibitem[\protect\citeauthoryear{{Applegate}, {Douglas}, {Gursel}, {Sussman}
  \& {Wisdom}}{{Applegate} et~al.}{1986}]{Applegate1986}
{Applegate} J.~H.,  {Douglas} M.~R.,  {Gursel} Y.,  {Sussman} G.~J.,   {Wisdom}
  J.,  1986, \mn@doi [\aj] {10.1086/114149}, \href
  {https://ui.adsabs.harvard.edu/abs/1986AJ.....92..176A} {92, 176}

\bibitem[\protect\citeauthoryear{Battin, of Aeronautics  \&
  Astronautics}{Battin et~al.}{1987}]{Battin.1987}
Battin R.,  of Aeronautics A.~I.,   Astronautics 1987, An Introduction to the
  Mathematics and Methods of Astrodynamics.
AIAA Textbook Series, American Institute of Aeronautics and Astronautics

\bibitem[\protect\citeauthoryear{Ba\`{u}, Urrutxua  \& Pelaez}{Ba\`{u}
  et~al.}{2014}]{Bauetal2014}
Ba\`{u} G.,  Urrutxua H.,   Pelaez J.,  2014, Adv. Astronaut. Sci., 152, 379

\bibitem[\protect\citeauthoryear{{Brouwer}}{{Brouwer}}{1937}]{B37}
{Brouwer} D.,  1937, \mn@doi [AJ] {10.1086/105423}, \href
  {http://adsabs.harvard.edu/abs/1937AJ.....46..149B} {46, 149}

\bibitem[\protect\citeauthoryear{Bulirsch \& Stoer}{Bulirsch \&
  Stoer}{1964}]{BulirschStoer1964}
Bulirsch R.,  Stoer J.,  1964, \mn@doi [Numerische Mathematik]
  {10.1007/BF01386092}, 6, 413

\bibitem[\protect\citeauthoryear{{Chambers}}{{Chambers}}{1999}]{C99}
{Chambers} J.~E.,  1999, \mn@doi [MNRAS] {10.1046/j.1365-8711.1999.02379.x},
  \href {http://adsabs.harvard.edu/abs/1999MNRAS.304..793C} {304, 793}

\bibitem[\protect\citeauthoryear{{Channell} \& {Scovel}}{{Channell} \&
  {Scovel}}{1990}]{chann90}
{Channell} P.~J.,  {Scovel} C.,  1990, \mn@doi [Nonlinearity]
  {10.1088/0951-7715/3/2/001}, 3, 231

\bibitem[\protect\citeauthoryear{{Danby}}{{Danby}}{1988}]{dan88}
{Danby} J.~M.~A.,  1988, {Fundamentals of celestial mechanics}, 2nd rev.~and
  enl. edn.
Willmann-Bell, Richmond, Va., U.S.A.

\bibitem[\protect\citeauthoryear{Deng, Wu  \& Liang}{Deng
  et~al.}{2020}]{Dengetal2020}
Deng C.,  Wu X.,   Liang E.,  2020, \mn@doi [Monthly Notices of the Royal
  Astronomical Society] {10.1093/mnras/staa1753}, 496, 2946

\bibitem[\protect\citeauthoryear{{Duncan}, {Levison}  \& {Lee}}{{Duncan}
  et~al.}{1998}]{DLL98}
{Duncan} M.~J.,  {Levison} H.~F.,   {Lee} M.~H.,  1998, \mn@doi [AJ]
  {10.1086/300541}, \href {http://adsabs.harvard.edu/abs/1998AJ....116.2067D}
  {116, 2067}

\bibitem[\protect\citeauthoryear{{Encke}}{{Encke}}{1854}]{E1854}
{Encke} J.,  1854, Berliner Astronomisches Jahrbuch f\"{u}r 1857, pp 319--397

\bibitem[\protect\citeauthoryear{{Everhart}}{{Everhart}}{1974}]{E74}
{Everhart} E.,  1974, \mn@doi [Celestial Mechanics] {10.1007/BF01261877}, \href
  {https://ui.adsabs.harvard.edu/abs/1974CeMec..10...35E} {10, 35}

\bibitem[\protect\citeauthoryear{{Everhart}}{{Everhart}}{1985}]{E85}
{Everhart} E.,  1985, in {Carusi} A.,  {Valsecchi} G.~B.,  eds, Dynamics of
  Comets: Their Origin and Evolution, Proceedings of IAU Colloq. 83, held in
  Rome, Italy, June 11-15, 1984. Edited by Andrea Carusi and Giovanni B.
  Valsecchi. Dordrecht: Reidel, Astrophysics and Space Science Library. Volume
  115, 1985, p.185. p.~185

\bibitem[\protect\citeauthoryear{Faou, Hairer  \& Pham}{Faou
  et~al.}{2004}]{FHP04}
Faou E.,  Hairer E.,   Pham T.-L.,  2004, \mn@doi [BIT Numerical Mathematics]
  {10.1007/s10543-004-5240-6}, 44, 699

\bibitem[\protect\citeauthoryear{González-Pinto, Pérez-Rodríguez  \&
  Montijano}{González-Pinto et~al.}{2001}]{Gonzalezpinto01}
González-Pinto S.,  Pérez-Rodríguez S.,   Montijano J.,  2001, \mn@doi
  [Computers & Mathematics with Applications]
  {https://doi.org/10.1016/S0898-1221(00)00335-7}, 41, 1009

\bibitem[\protect\citeauthoryear{Grazier, Newman, Hyman, Sharp  \&
  Goldstein}{Grazier et~al.}{2005}]{Graz2005}
Grazier K.~R.,  Newman W.~I.,  Hyman J.~M.,  Sharp P.~W.,   Goldstein D.~J.,
  2005, in May R.,  Roberts A.~J.,  eds,  Vol. 46, Proc. of 12th Computational
  Techniques and Applications Conference CTAC-2004. pp C786--C804

\bibitem[\protect\citeauthoryear{{Hairer}, {Lubich}  \& {Wanner}}{{Hairer}
  et~al.}{2006}]{hair06}
{Hairer} E.,  {Lubich} C.,   {Wanner} G.,  2006, {Geometrical Numerical
  Integration}, 2nd edn.
Springer Verlag, Berlin

\bibitem[\protect\citeauthoryear{{Hairer}, {McLachlan}  \&
  {Razakarivony}}{{Hairer} et~al.}{2008}]{HMR07}
{Hairer} E.,  {McLachlan} R.,   {Razakarivony} A.,  2008, BIT, 48, 231

\bibitem[\protect\citeauthoryear{{Hayes}}{{Hayes}}{2007}]{Hayes2007}
{Hayes} W.~B.,  2007, \mn@doi [Nature Physics] {10.1038/nphys728}, \href
  {https://ui.adsabs.harvard.edu/abs/2007NatPh...3..689H} {3, 689}

\bibitem[\protect\citeauthoryear{{Hernandez}}{{Hernandez}}{2019a}]{H19}
{Hernandez} D.~M.,  2019a, \mn@doi [\mnras] {10.1093/mnras/stz884}, \href
  {https://ui.adsabs.harvard.edu/abs/2019MNRAS.486.5231H} {486, 5231}

\bibitem[\protect\citeauthoryear{{Hernandez}}{{Hernandez}}{2019b}]{H19b}
{Hernandez} D.~M.,  2019b, \mn@doi [\mnras] {10.1093/mnras/stz2662}, \href
  {https://ui.adsabs.harvard.edu/abs/2019MNRAS.490.4175H} {490, 4175}

\bibitem[\protect\citeauthoryear{{Hernandez} \& {Bertschinger}}{{Hernandez} \&
  {Bertschinger}}{2018}]{HB18}
{Hernandez} D.~M.,  {Bertschinger} E.,  2018, \mn@doi [MNRAS]
  {10.1093/mnras/sty184}, \href
  {http://adsabs.harvard.edu/abs/2018MNRAS.475.5570H} {475, 5570}

\bibitem[\protect\citeauthoryear{{Hernandez} \& {Dehnen}}{{Hernandez} \&
  {Dehnen}}{2017}]{HD17}
{Hernandez} D.~M.,  {Dehnen} W.,  2017, \mn@doi [\mnras]
  {10.1093/mnras/stx547}, \href
  {http://adsabs.harvard.edu/abs/2017MNRAS.468.2614H} {468, 2614}

\bibitem[\protect\citeauthoryear{{Hernandez}, {Hadden}  \&
  {Makino}}{{Hernandez} et~al.}{2020}]{Hetal2020}
{Hernandez} D.~M.,  {Hadden} S.,   {Makino} J.,  2020, \mn@doi [\mnras]
  {10.1093/mnras/staa388}, \href
  {https://ui.adsabs.harvard.edu/abs/2020MNRAS.493.1913H} {493, 1913}

\bibitem[\protect\citeauthoryear{Kahan}{Kahan}{1965}]{Kahan1965}
Kahan W.,  1965, \mn@doi [Commun. ACM] {10.1145/363707.363723}, 8, 40

\bibitem[\protect\citeauthoryear{{Leimkuhler} \& {Reich}}{{Leimkuhler} \&
  {Reich}}{2004}]{leim04}
{Leimkuhler} B.,  {Reich} S.,  2004, {Simulating Hamiltonian Dynamics}.
Cambridge University Press

\bibitem[\protect\citeauthoryear{{Makino}}{{Makino}}{1991}]{Makino1991}
{Makino} J.,  1991, \mn@doi [\apj] {10.1086/169751}, \href
  {https://ui.adsabs.harvard.edu/abs/1991ApJ...369..200M} {369, 200}

\bibitem[\protect\citeauthoryear{{Press}, {Teukolsky}, {Vetterling}  \&
  {Flannery}}{{Press} et~al.}{2002}]{press02}
{Press} W.~H.,  {Teukolsky} S.~A.,  {Vetterling} W.~T.,   {Flannery} B.~P.,
  2002, {Numerical recipes in C++ : the art of scientific computing}

\bibitem[\protect\citeauthoryear{{Quinlan} \& {Tremaine}}{{Quinlan} \&
  {Tremaine}}{1990}]{QT90}
{Quinlan} G.~D.,  {Tremaine} S.,  1990, \mn@doi [\aj] {10.1086/115629}, \href
  {https://ui.adsabs.harvard.edu/abs/1990AJ....100.1694Q} {100, 1694}

\bibitem[\protect\citeauthoryear{{Rein} \& {Spiegel}}{{Rein} \&
  {Spiegel}}{2015}]{RS15}
{Rein} H.,  {Spiegel} D.~S.,  2015, \mn@doi [MNRAS] {10.1093/mnras/stu2164},
  \href {http://adsabs.harvard.edu/abs/2015MNRAS.446.1424R} {446, 1424}

\bibitem[\protect\citeauthoryear{{Rein} \& {Tamayo}}{{Rein} \&
  {Tamayo}}{2015}]{RT15}
{Rein} H.,  {Tamayo} D.,  2015, \mn@doi [MNRAS] {10.1093/mnras/stv1257}, \href
  {http://adsabs.harvard.edu/abs/2015MNRAS.452..376R} {452, 376}

\bibitem[\protect\citeauthoryear{{Rein} et~al.,}{{Rein}
  et~al.}{2019a}]{Reinetal2019}
{Rein} H.,  et~al., 2019a, \mn@doi [\mnras] {10.1093/mnras/stz769}, \href
  {https://ui.adsabs.harvard.edu/abs/2019MNRAS.485.5490R} {485, 5490}

\bibitem[\protect\citeauthoryear{{Rein}, {Tamayo}  \& {Brown}}{{Rein}
  et~al.}{2019b}]{Reinetal2019b}
{Rein} H.,  {Tamayo} D.,   {Brown} G.,  2019b, \mn@doi [\mnras]
  {10.1093/mnras/stz2503}, \href
  {https://ui.adsabs.harvard.edu/abs/2019MNRAS.489.4632R} {489, 4632}

\bibitem[\protect\citeauthoryear{{Rein}, {Brown}  \& {Tamayo}}{{Rein}
  et~al.}{2019c}]{Rein2019}
{Rein} H.,  {Brown} G.,   {Tamayo} D.,  2019c, \mn@doi [MNRAS]
  {10.1093/mnras/stz2942}, \href
  {https://ui.adsabs.harvard.edu/abs/2019MNRAS.490.5122R} {490, 5122}

\bibitem[\protect\citeauthoryear{{Ruth}}{{Ruth}}{1983}]{Ruth1983}
{Ruth} R.~D.,  1983, \mn@doi [IEEE Transactions on Nuclear Science]
  {10.1109/TNS.1983.4332919}, \href
  {https://ui.adsabs.harvard.edu/abs/1983ITNS...30.2669R} {30, 2669}

\bibitem[\protect\citeauthoryear{Stiefel \& Scheifele}{Stiefel \&
  Scheifele}{1971}]{SS71}
Stiefel E.,  Scheifele G.,  1971.

\bibitem[\protect\citeauthoryear{{Tamayo}, {Rein}, {Shi}  \&
  {Hernandez}}{{Tamayo} et~al.}{2020}]{Tamayoetal2019}
{Tamayo} D.,  {Rein} H.,  {Shi} P.,   {Hernandez} D.~M.,  2020, \mn@doi
  [\mnras] {10.1093/mnras/stz2870}, \href
  {https://ui.adsabs.harvard.edu/abs/2020MNRAS.491.2885T} {491, 2885}

\bibitem[\protect\citeauthoryear{{Tsang}, {Galley}, {Stein}  \&
  {Turner}}{{Tsang} et~al.}{2015}]{TGS2015}
{Tsang} D.,  {Galley} C.~R.,  {Stein} L.~C.,   {Turner} A.,  2015, \mn@doi
  [\apjl] {10.1088/2041-8205/809/1/L9}, \href
  {https://ui.adsabs.harvard.edu/abs/2015ApJ...809L...9T} {809, L9}

\bibitem[\protect\citeauthoryear{{Wisdom}}{{Wisdom}}{1982}]{Wisdom82}
{Wisdom} J.,  1982, \mn@doi [\aj] {10.1086/113132}, \href
  {https://ui.adsabs.harvard.edu/abs/1982AJ.....87..577W} {87, 577}

\bibitem[\protect\citeauthoryear{{Wisdom}}{{Wisdom}}{1983}]{W83}
{Wisdom} J.,  1983, \mn@doi [Icarus] {10.1016/0019-1035(83)90127-6}, \href
  {http://adsabs.harvard.edu/abs/1983Icar...56...51W} {56, 51}

\bibitem[\protect\citeauthoryear{{Wisdom}}{{Wisdom}}{2018}]{Wisdom2018}
{Wisdom} J.,  2018, \mn@doi [\mnras] {10.1093/mnras/stx2906}, \href
  {https://ui.adsabs.harvard.edu/abs/2018MNRAS.474.3273W} {474, 3273}

\bibitem[\protect\citeauthoryear{{Wisdom} \& {Hernandez}}{{Wisdom} \&
  {Hernandez}}{2015}]{WH15}
{Wisdom} J.,  {Hernandez} D.~M.,  2015, \mn@doi [MNRAS]
  {10.1093/mnras/stv1862}, \href
  {http://adsabs.harvard.edu/abs/2015MNRAS.453.3015W} {453, 3015}

\bibitem[\protect\citeauthoryear{{Wisdom} \& {Holman}}{{Wisdom} \&
  {Holman}}{1991}]{WH91}
{Wisdom} J.,  {Holman} M.,  1991, \mn@doi [AJ] {10.1086/115978}, 102, 1528

\makeatother
\end{thebibliography}

\end{document}